\documentclass[12pt,preprintnumbers,superscriptaddress,amsmath,amssymb,nofootinbib]{revtex4}
\usepackage{graphicx}
\usepackage{dcolumn}
\usepackage{bm}
\usepackage{amssymb}
\usepackage{amsmath}
\usepackage{epsfig}    
\usepackage{color}
\usepackage{slashed}
\usepackage{hhline}
\def\be{\begin{equation}}
\def\ee{\end{equation}}
\newcommand{\bea}{\begin{eqnarray}}
\newcommand{\eea}{\end{eqnarray}}
\newcommand{\nn}{\nonumber}

\numberwithin{equation}{section}

\begin{document}

\title{A simple model for explaining muon-related anomalies and dark matter}

\author{Cheng-Wei Chiang}
\email{chengwei@phys.ntu.edu.tw}
\affiliation{Department of Physics, National Taiwan University, Taipei 10617, Taiwan}
\affiliation{Institute of Physics, Academia Sinica, Taipei 11529, Taiwan}
\affiliation{Physics Division, National Center for Theoretical Sciences, Hsinchu 30013, Taiwan}


\author{Hiroshi Okada}
\email{hiroshi.okada@apctp.org}
\affiliation{Asia Pacific Center for Theoretical Physics (APCTP) - Headquarters
San 31, Hyoja-dong, Nam-gu, Pohang 790-784, Korea}

\date{\today}

\begin{abstract}
We propose a model to explain several muon-related experimental anomalies and the abundance of dark matter.  We introduce an vector-like exotic lepton that form an iso-doublet and three right-handed Majorana fermions as an iso-singlet.  A real/complex scalar field is added as a dark matter candidate.  We impose a global $U(1)_\mu$ symmetry under which fields associated with the SM muon are charged.  To stabilize the dark matter, we impose a $Z_2$ (or $Z_3$) symmetry under which the exotic lepton doublet and the real (or complex) scalar field are charged.  We find that the model can simultaneously explain the muon anomalous magnetic dipole moment and the dark matter relic density in no conflict with any lepton flavor-violating/conserving observables,
with some details depending upon whether the scalar field is real or complex.  Besides, we extend the framework to the quark sector in a way similar to the lepton sector, and find that the recent anomalies associated with the $b \to s \mu^+ \mu^-$ transition can also be accommodated while satisfying constraints such as the $B_{s(d)}\to \mu^+\mu^-$ decays and neutral meson mixings.
\end{abstract}
\maketitle

\newpage

\section{Introduction}

In search of new physics, most results from the Large Hadron Collider (LHC) at the energy frontier are consistent with the Standard Model (SM) predictions and only push the existence of new particles to higher scales.  On the other hand, we have encountered over the past few years a few observables in low-energy flavor physics that show evidence of deviations from the SM expectations.  Interestingly, many of these processes involve the muon.

A long-standing puzzle is the muon anomalous magnetic dipole moment, $a_\mu$ or $(g-2)_\mu$.  With advances in theory and inputs from various experiments, $a_\mu$ has been calculated to a high precision.  In comparison with experimental data, we observe a discrepancy at the 3.3$\sigma$ level:\
\footnote{There are other analyses giving slightly different estimates of the discrepancy.  For example, Ref.~\cite{Benayoun:2011mm} gives $\Delta a_\mu= (33.5\pm 8.2)\times 10^{-10}$ showing a discrepancy at the 4.1$\sigma$ level, while Ref.~\cite{fermi-lab} quotes $\Delta a_\mu= 288(63)(49)\times 10^{-11}$ indicating a 3.5$\sigma$ deviation.  In our numerical analysis, we use the result given by Ref.~\cite{Hagiwara:2011af}.}
$\Delta a_\mu=(26.1\pm 8.0)\times 10^{-10}$~\cite{Hagiwara:2011af}.

Recently, some evidence of deviations seemed to occur in decays involving the $b \to s \mu^+\mu^-$ transition, such as the binned angular distribution of the $B\to K^{*} \mu^+ \mu^-$ decay~\cite{Aaij:2013qta, Aaij:2015oid, Abdesselam:2016llu, Wehle:2016yoi} and the decay rate deficit of the $B_s \to K^{(*)}  \mu^+ \mu^-$ and $B_s \to \phi  \mu^+ \mu^-$ decays~\cite{Aaij:2013aln, Aaij:2015esa}.  More recently, the LHCb Collaboration reported anomalies in a set of related observables, $R_K \equiv BR(B \to K \mu^+ \mu^-)/BR(B \to K e^+ e^-)$ and $R_{K^*} \equiv BR(B \to K^{*} \mu^+ \mu^-)/BR(B \to K^{*} e^+ e^-)$.  The former was found to be $0.745^{+0.090}_{-0.074}\pm0.036$ for the dilepton invariant mass-squared range \,$q^2\in(1,6)~{\rm GeV}^2$~\cite{Aaij:2014ora}, showing a $2.6\sigma$ deviation.  The latter was determined in two dilepton $q^2$ bins:
\begin{align*}
R_{K^*}
=
\begin{cases}
0.66^{+0.11}_{-0.07} \pm 0.03 & \mbox{for } q^2 \in [0.045 , 1.1]~\mbox{GeV}^2 ~,
\\
0.69^{+0.11}_{-0.07} \pm 0.05 & \mbox{for } q^2 \in [1.1 , 6]~\mbox{GeV}^2 ~.
\end{cases}
\end{align*}
These two observables point to lepton non-universality in the $B \to K^{(*)} \ell^+ \ell^-$ ($\ell = e, \mu$) decays.  Depending on scenarios~\cite{Descotes-Genon:2015uva}, global fits to the data reveal deviations in the Wilson coefficients in the related weak decay Hamiltonian, most notably in $C_9$ associated with the operator $O_9^\ell = \frac{\alpha}{4\pi} (\bar s_L \gamma^\mu b_L) (\overline{\ell}\gamma_\mu \ell)$.

Motivated by the above-mentioned flavor anomalies, we propose a simple model with interactions specific to the muon.  In addition to the SM particles, we introduce an exotic vector-like lepton doublet and three right-handed neutrinos for each SM lepton family and an inert scalar boson that can be either real or complex.  A global $U(1)_\mu \times Z_N$ symmetry is imposed on the model, with the muon-related fields (including the left-handed and right-handed muons and the associated exotic muon) charged under the $U(1)_\mu$ and the exotic lepton doublet and the inert scalar carrying nontrivial charges under the $Z_N$.  Here $N = 2$ if the scalar field is real and $N = 3$ as a minimal choice if it is complex.  With the imposed $Z_2$ or $Z_3$ symmetry, we make a connection to the observed dark matter (DM) relic abundance in the Universe~\cite{Ade:2013zuv}, with the inert scalar particle serving as a bosonic weakly-interacting massive particle (WIMP) candidate.  Due to the muon-specific interactions, there is a strong correlation between $a_\mu$ and DM parameters.  By extending the model to the quark sector in a way analogous to the lepton sector except for no additional global $U(1)$ symmetry, we find that the $b \to s \ell^+\ell^-$ anomalies can be accommodated without conflict with various constraints such as the $B_{s(d)}\to \mu^+\mu^-$ decays, neutral meson mixings, and lepton flavor-violating (LFV) observables. {We note that the new global symmetry plays a crucial role in model construction and renders a minimal framework for accommodating all the above-mentioned muon-related anomalies.}

This paper is organized as follows.  In Section~\ref{sec:model}, we describe the proposed model with the extended lepton sector and the inert scalar, and show the contributions to $a_\mu$, and DM relic density.
{We also consider the constraint coming from the $Z \to \mu^+ \mu^-$ decay at the one-loop level.
Section~\ref{sec:Zll} is devoted to the discussion about lepton flavor-violating processes at the one-loop level when we extend the model to have three exotic lepton families.}
In Section~\ref{sec:extension}, we extend the model to the quark sector to include the exotic quark fields and formulate the effective weak Hamiltonian for the $b \to s \ell^+ \ell^-$ transitions.  It is then used to explain the above-mentioned $B$ anomalies subject to various constraints.  Section~\ref{sec:fit} combines the analysis in the previous two sections and shows the result of a global fit.  Section~\ref{sec:summary} summarizes our findings.

\section{Model Setup \label{sec:model}}

In this section, we concentrate on the lepton and scalar sectors (all assumed to be colorless) of the model, and will discuss the quark sector in the next section.  In addition to the SM $SU(3)_C \times SU(2)_L \times U(1)_Y$ gauge group, we impose on our model an additional global $U(1)_\mu \times Z_2$ or $U(1)_\mu \times Z_3$ symmetry\
\footnote{We note that the $Z_3$ symmetry can be generalized to $Z_N$ with $N > 3$.  In this case, the $Z_N$ charge of $L'_\mu$ is $\tilde\omega \equiv e^{2\pi i / N}$ and that of $S$ is $\tilde\omega^{-1}$.  What this affects is the allowed interactions of the $S$ field in the scalar potential.}, depending upon the choice of a new inert scalar field.  For each distinct SM lepton family, we introduce corresponding an $SU(2)_L$-doublet vector-like fermion $L'_\mu = [N'_\mu,E'_\mu]^T$ and three $SU(2)_L$-singlet right-handed Majorana fermions ${N_{R_i}}$ ($i = e, \mu, \tau$).  These exotic leptons are assumed to be heavier than their SM counterparts.  Moreover, the fermions in the second families of SM and exotic leptons carry an $U(1)_{\mu}$ charge, denoted by $Q_\mu$.  This serves the purpose of evading the $\mu\to e \gamma$ constraint, as to be discussed later.  We also introduce an $SU(2)_L$-singlet scalar boson $S$, which does not carry any gauge or $U(1)_{\mu}$ charge.  We will consider both possibilities of $S$ being real or complex.  In this set up, only the exotic lepton doublet $L'_{\mu}$ and the inert scalar boson $S$ carry nontrivial $Z_2$ or $Z_3$ charges.  In the case of a real $S$, these fields all have the $Z_2$ charge of $-1$.  In the case of a complex $S$, $L'_{\mu}$ and $S$ have respective charges of $\omega \equiv e^{2\pi i/3}$ and $\omega^2$ under the $Z_3$ symmetry.  This provides a mechanism to prevent mixing between the SM fields and the exotic fields as well as to maintain the stability of the DM candidate $S$\
\footnote{The neutral component of $L'_\mu$ cannot be DM candidate, as they would be ruled out by direct detection via the $Z$ boson portal.}.
The field contents and their charge assignments are summarized in Table~\ref{tab:1}.

\begin{widetext}
\begin{center} 
\begin{table}[tb]
\begin{tabular}{|c||c|c|c|c|c|c|c||c|c|}\hline
&\multicolumn{7}{c||}{Lepton Fields} & \multicolumn{2}{c|}{Scalar Fields} \\
\hline
& ~$L_{L_{e,\tau}}$~& ~$L_{L_\mu}$~ & ~$\ell_{R_{e,\tau}}$ & ~$\ell_{R_\mu}$~ & ~$L'_{\mu}$~ &~ $N_{R_{e,\tau}}$~ & ~$N_{R_\mu}$~ & ~$H$~  & ~$S$~ \\
\hline 
$SU(2)_L$ & $\bm{2}$ & $\bm{2}$  & $\bm{1}$ & $\bm{1}$ & $\bm{2}$& $\bm{1}$ & $\bm{1}$  & $\bm{2}$ & $\bm{1}$ \\\hline 
$U(1)_Y$ & $-1/2$& $-1/2$ & $-1$& $-1$  & $-1/2$  & $0$& $0$ & $1/2$ & $0$   \\\hline
$U(1)_{\mu}$ & $0$ & $Q_\mu$   & $0$ & $Q_\mu$  & $Q_\mu$ & $0$ & $Q_\mu$ & $0$& $0$ \\\hline
$Z_2$ & $1$ & $1$ & $1$ & $1$   & $-1$ & $1$& $1$ & $1$ & $-1$ \\
$Z_3$ & $1$ & $1$ & $1$ & $1$  & $\omega$ & $1$& $1$ & $1$ & $\omega^2$ \\\hline
\end{tabular}
\caption{Contents of colorless fermion and scalar fields in the model, with their charge assignments under the SM $SU(2)_L\times U(1)_Y$ and global $U(1)_\mu\times Z_{2,3}$ symmetries.  The row of $Z_2$ ($Z_3$) symmetry is for the scenario when $S$ is a real (complex) field.}
\label{tab:1}
\end{table}
\end{center}
\end{widetext}

In the most general renormalizable Lagrangian consistent with the symmetries of the model, the lepton sector and the Higgs potential are given respectively by 
\begin{align}
-\mathcal{L}_L
=& 
y_{\ell_i} \bar L_i H P_R \ell_i 
+ f_\mu \bar L_\mu P_R L'_\mu S 
+ y_{N_{i}} \bar L_i \tilde H P_R N_{i} 
+ y_{N_{e\tau}} \bar L_e \tilde H P_R N_{\tau} + y_{N_{\tau e}} \bar L_\tau \tilde H P_R N_{e} 
\nn\\
&
+ M_\mu \bar L'_\mu P_R L'_\mu 
+ M_{N_{ij}} \bar N_{i} P_RN_{j} 
+ {\rm H.c.}
~, 
\label{eq:lag}\\
V=&
\mu_H^2 |H|^2 + \mu_S^2 |S|^2 
+ \lambda_H |H|^4 + \lambda_S |S|^4 + \lambda_{HS} |H|^2 |S|^2 + \mu (S^3  + {S^*}^3)
~, 
\label{Eq:pot} 
\end{align}
where $i,j \in \{ e,\mu,\tau \}$ 
are to be summed over when repeated, the charged-lepton mass is assumed diagonal without loss of generality, and $\tilde H\equiv (i\sigma_2)H^*$ with $\sigma_2$ being the second Pauli matrix. 
Note here that $M_N$ is softly broken under the $U(1)_\mu$ symmetry due to generate the observed neutrino masses and mixings.
{It suggests that the charged leptons can be regarded as in their mass eigenstates from the beginning.}
The Higgs potential given above is the one with a $Z_3$ symmetry.  The one with a $Z_2$ symmetry can be readily obtained by taking $\mu = 0$.
As in the SM, the first term of the Yukawa Lagrangian, Eq.~\eqref{eq:lag}, provides mass for SM charged leptons when $H$ develops a nonzero vacuum expectation value (VEV), $\langle H\rangle\equiv v/\sqrt2$.
In particular, the $f_\mu$ term contributes to the muon anomalous magnetic dipole moment at the one-loop level.  
{With the $\mu$ terms, we then have $SS\to SL'$ (s-channel) or $SS\to SH$ (t,u-channels) scattering processes whose semi-annihilation modes can modify the phenomenology such as the DM relic density.  Such effects can be used to discern between the $Z_2$ and $Z_3$ scenarios.}

\subsection{Neutrino sector}

Mass of the active neutrinos can be induced via the canonical seesaw mechanism, and the mass matrix is given by
 \begin{align}
&({\cal M}_{\nu})_{\alpha\beta}  \approx
\sum_{i,j=e,\mu,\tau} (m_D)_{\alpha i} \left( M_{N}^{-1} \right)_{ij} \left( {m^T_D} \right)_{j\beta} ,
\end{align}
where $m_D\equiv y_N v / \sqrt2$ and $\alpha, \beta \in \{ e, \tau \}$.
The mass matrix ${\cal M}_{\nu}$ is then diagonalized as $D_\nu\equiv U_{PMNS} {\cal M}_{\nu} U_{PMNS}^T$, where $U_{PMNS}$ and $D_\nu$ can be determined using the current neutrino oscillation data~\cite{Gonzalez-Garcia:2014bfa}.

Without loss of generality, we work in the basis where all the coefficients in the scalar potential \eqref{Eq:pot} are real, and parameterize the SM scalar doublet as
\begin{align}
&H =
\begin{pmatrix}
w^+\\
\frac{1}{\sqrt2} (v+h+iz)
\end{pmatrix}
\qquad \mbox{with}~v = 246~{\rm GeV}
~,
\label{eq:hp-cond}
\end{align}
where $w^+$ and $z$ are to be absorbed by the SM $W$ and $Z$ bosons, respectively.  Moreover, we assume that the $S$ field does not develop a nonzero VEV.
To stabilize the scalar potential and to have a global minimum given by Eq.~(\ref{eq:hp-cond}), the quartic couplings should satisfy the following conditions~\cite{Barbieri:2006dq, Belanger:2012vp}:
\begin{align}
{0<\lambda_S,\lambda_H, \quad   \lambda_{HS}^2<4{\lambda_H \lambda_S} \quad\mbox{and}\quad 
\frac98\frac{\mu^2}{\lambda_S}+\frac{\lambda_{HS}}{\lambda_H}\mu_H^2<\mu_S^2.}
\label{eq:stability}
\end{align}

\subsection{Muon anomalous magnetic dipole moment \label{damu}}

The interaction relevant to ${\Delta a_\mu}$ is
\begin{align}
f_\mu \bar\ell_\mu P_R E'_\mu S + {\rm H.c.}\ .
\end{align}
With $S$ of mass $m_S$ and $E'_\mu$ of mass $M_{\mu'}$ running in the loop, we obtain
\begin{align}
\Delta a_\mu 
=
\frac{|f_{\mu}|^2} {{8} \pi^2} 
{\int_0^1 dx \frac{m_\mu^2 x^2(1-x)}{x(x-1) + M_{\mu'}^2 x + (1-x) m_S^2}} ~.
\label{eq:damu-res}
\end{align}
To explain the current 3.3$\sigma$ deviation~\cite{Hagiwara:2011af}
\begin{align}
\Delta a_\mu = (26.1\pm8.0)\times10^{-10} ~,
\label{eq:damu}
\end{align}
the model has three degrees of freedom: $|f_\mu|$, $m_S$, and $M_{\mu'}$
\footnote{{For a comprehensive review on new physics models for the muon $g-2$ anomaly as well as lepton flavor violation, please see Ref.~\cite{Lindner:2016bgg}.}}.

\subsection{Bosonic dark matter candidate}

Stabilized by the $Z_2$ or $Z_3$ symmetry, the $S$ boson serves as a DM candidate.
We first discuss the bounds coming from the spin independent scattering cross section reported by several direct detection experiments such as LUX~\cite{Akerib:2016vxi}, XENON1T~\cite{Aprile:2017iyp}, and PandaX-II~\cite{Cui:2017nnn}, as in our model there is a Higgs portal contribution.  We have checked that as long as ${\lambda_{HS}} \lesssim {\cal O}$(0.01), there is no constraint from direction detection.  Therefore, we assume in this work that these quartic couplings are sufficiently small but satisfy Eq.~\eqref{eq:stability}.

The relevant terms for the relic density of the $S$ boson are
\begin{align}
{f_\mu}(\bar\ell_\mu P_R E'_\mu + \bar\nu_\mu P_R N'_\mu) S +{\rm H.c.}
~,
\end{align}
where the other $f$ terms are assumed to be negligible in comparison with $f_{\mu}$, as a larger value of  $f_{\mu}$ is required to obtain a sizable $\Delta a_\mu$.  Such interactions will lead to pair annihilation of the $S$ bosons in the SM muons and muon neutrinos.
To explicitly evaluate the relic abundance of $S$, one has to specify whether the $S$ field is real or complex.
For a real $S$ we have both $t$- and $u$-channel annihilation processes that lead to a more suppressed $d$-wave cross section, while for a complex $S$, on the other hand, there is only the $t$ channel that leads to a $p$-wave dominant cross section.
The cross sections for the two scenarios are approximately given by~{\cite{Giacchino:2013bta}}
\begin{align}
&(\sigma v _{\rm rel})({2S}\to \mu\bar\mu (\nu_\mu\bar\nu_\mu))
\approx 
\frac{ |f_{\mu}|^4}{60 \pi} 
\frac{m_S^6}{(m_S^2+M^2_{\mu'})^4} v_{\rm rel}^4, \quad ({\rm Real \ S}) 
\label{eq:anni-rl}
~, \\
&(\sigma v _{\rm rel})({SS^*}\to \mu\bar\mu (\nu_\mu\bar\nu_\mu))
\approx 
\frac{ |f_{\mu}|^4}{96 \pi} 
\frac{m_S^2}{(m_S^2+M^2_{\mu'})^2} v_{\rm rel}^2, \quad ({\rm Complex \ S}) 
\label{eq:anni-cplx}
~,
\end{align}
in the limit of massless final-state leptons.  Here the approximate formulas are obtained by expanding the cross sections in powers of the relative velocity $v_{\rm rel}$: $\sigma v _{\rm rel}\approx a_{\rm eff} + b_{\rm eff} v^2_{\rm rel} + d_{\rm eff} v^4_{\rm rel}$.
The resulting relic densities of the two scenarios are found to be
\begin{align}
&\Omega h^2\approx 
\frac{5.35\times 10^7 x_f^3}{\sqrt{g_*(x_f)} M_{\rm PL} d_{\rm eff}} \quad ({\rm Real \ S})  \label{eq:relic-rl}
~,\\
&\Omega h^2\approx 
\frac{1.78\times 10^8 x_f^2}{\sqrt{g_*(x_f)} M_{\rm PL} b_{\rm eff}} \quad ({\rm Complex \ S})  \label{eq:relic-cplx}
~,
\end{align}
respectively, where the present relic density is $0.1199 \pm 0.0054$ at the 2$\sigma$ confidential level (CL)~\cite{Ade:2013zuv}, $g_*(x_f\approx 25)\approx100$ counts the degrees of freedom for relativistic 
particles, and $M_{\rm PL}\approx 1.22\times 10^{19}$~GeV is the Planck mass.

Taking the central value of the relic density as an explicit example, the above formulas can be simplified to give: 
\begin{align}
&|f_\mu|\approx 0.057\times 
\frac{m_S^2+M_{\mu'}^2}{m_S^{3/2} \cdot{\rm GeV}^{1/2}}
\quad ({\rm Real \ S}) \label{eq:bs-relic-rl} ~, \\
&|f_\mu|\approx 0.033\times 
\sqrt{\frac{m_S^2+M_{\mu'}^2}{m_S \cdot {\rm GeV} }}
\quad ({\rm Complex \ S}) ~, \label{eq:bs-relic-cplx}
\end{align}
for which one still has to impose the perturbativity upper bound of $\sqrt{4\pi}$.
It is then straightforward to search for viable parameter space in the {$(m_S,M_{\mu'})$} plane by combining Eq.~(\ref{eq:bs-relic-rl}) or (\ref{eq:bs-relic-cplx}) with Eq.~(\ref{eq:damu-res}).

{
\subsection{ Lepton Flavor-Conserving  $Z$ Boson Decay}
Here, we consider the flavor conserved $Z$ boson decay $Z\to \mu^+ \mu^-$, and its decay rate is given by
\begin{align}
\Gamma(Z\to \mu^+ \mu^-)=83.95\pm0.44\ {\rm MeV},\label{eq:z2mu-exp}
\end{align}
where we have used the relaxed measurement at L3~\cite{pdg}.
Our decay rate is found as
\begin{align}
\text{BR}(Z\to\mu^+ \mu^-)
&=
\frac{G_F m_Z^3}{3072\sqrt2 \pi^5}
\left(s_w^2-\frac12\right)^2
\left( |f_{\mu}|^4 (1-2s_w^2)^2 \left| F_2(\mu',S)+F_3(\mu',S) \right|^2\right.\\
&\left.
+32\pi^2 |f_\mu|^2 (1-2s_w^2)^2 \left[ F_2(\mu',S)+F_3(\mu',S)\right]
+256 \pi^4(1-4s_w^2+8s_w^4)
\right)
~\nn,
\label{eq:Z2mu}
\end{align}
where
\begin{align*}
F_2(a,b) &=\int_0^1dx(1-x)\ln\left[xM_a^2 +  (1-x) m_b^2 \right] 
~,
\\
F_3(a,b) &=\int_0^1dx\int_0^{1-x}dy\frac{2xy m_Z^2+(M_a^2-m_b^2)(1-x-y)-\Delta\ln\Delta}{\Delta} 
~,
\end{align*} 
and $\Delta\equiv -xy m_Z^2 +(x+y)(M_a^2=m_b^2)+m_b^2$, $G_F\approx 1.17\times10^{-5}$ GeV$^{-2}$ is the Fermi decay constant~\cite{Mohr:2015ccw},
$m_Z\approx91$ GeV is the neutral vector  boson in the SM~\cite{pdg}, $s_w(\equiv \sin\theta_w)\approx0.22$
is Weinberg angle~\cite{pdg},
It gives a constraint from the fact that the above equation should be within the range of Eq.~(\ref{eq:z2mu-exp}).

}

\subsection{Global analysis \label{sec:fit}}

To perform a global analysis of the model, we require that both $\Delta a_\mu$ and the DM relic density fall within the $1\sigma$ range of the measured data and that the LFV processes satisfy their respective upper bounds quoted above.
In addition, we restrict ourselves to the {mass ranges}
\begin{align}
& m_S\in [1,400]\ {\rm GeV} 
~, \quad 
M_{\mu'}\in [100, 500]\ {\rm GeV}
~, \quad 
M_{\ell'_i}\in [1.2 M_X ,2\ {\rm TeV}]
~,
\end{align}
where $1.2 m_S \le M_{\mu'}$ is imposed to prevent the possibility of co-annihilation as well as the stability of $S$.

\begin{figure}[tb]
\begin{center}
\includegraphics[width=80mm]{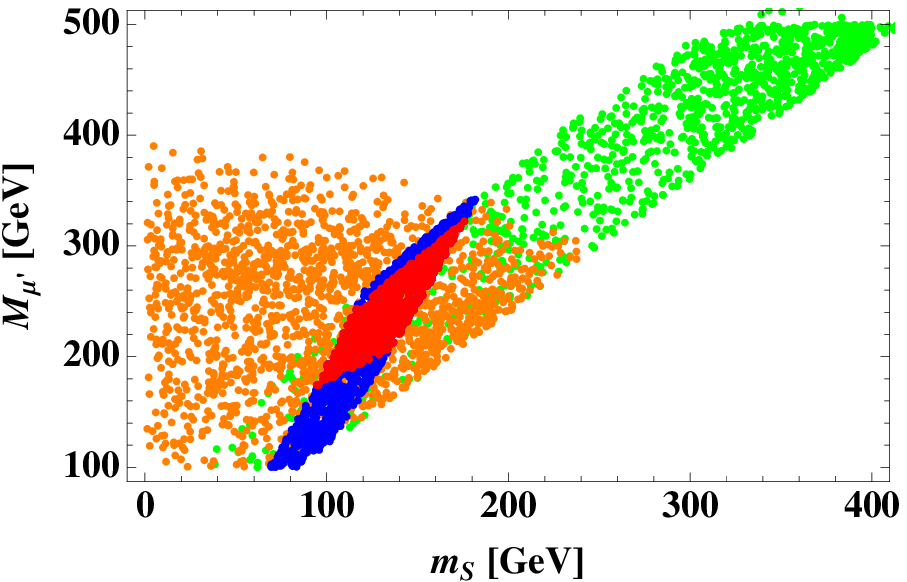}
\includegraphics[width=80mm]{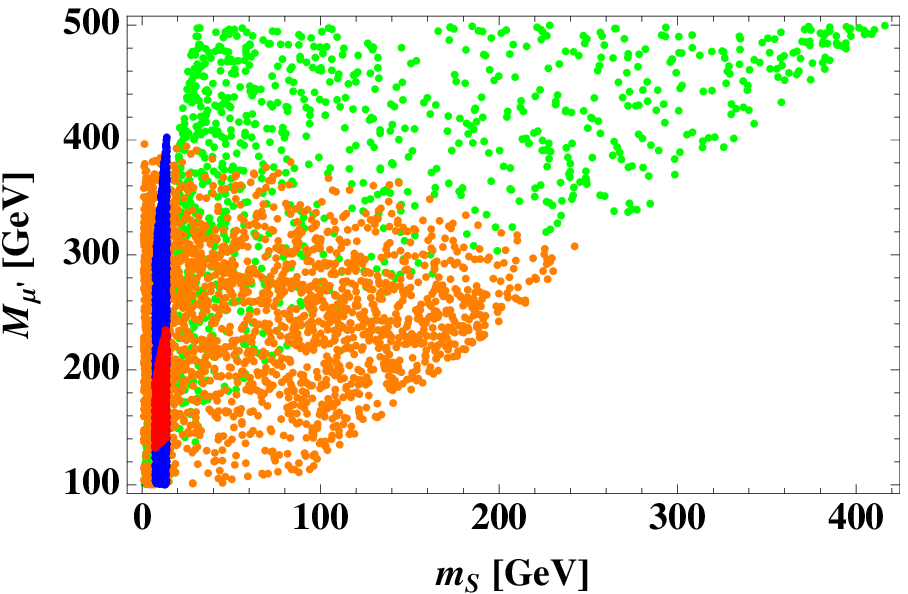}
\caption{Allowed parameter space in the $(m_S,M_{\mu'})$ plane,
{where the left (right) plot is for the scenario where $S$ is a real (complex) scalar boson.}
  While satisfying the LFV constraints, the orange (green) dots further satisfy the current $\Delta a_\mu$ (the current $\Omega h^2$) value at the $1\sigma$ level.
{The red (blue) dots are allowed by all of the constraints (except for BR$(Z\to\mu^+ \mu^-)$). }
}
\label{fig:dm-ME}
\end{center}
\end{figure}

Fig.~\ref{fig:dm-ME} shows the allowed parameter space in the $(m_S,M_{\mu'})$ plane by scanning all the other parameters.
The left (right) plot is for the scenario where $S$ is a real (complex) scalar boson.  In both plots and on top of the LFV constraints, the orange (green) dots further satisfy the current $\Delta a_\mu$ (the current $\Omega h^2$) value at the $1\sigma$ level.  In these scatter dots, only {the red (blue) dots are allowed by all of the constraints (except for BR$(Z\to\mu^+ \mu^-)$).
The left plot shows that the real $S$ scenario favors the parameter space of 90~(70)~GeV~$\lesssim m_S\lesssim$~180~(185)~GeV and 160~(100)~GeV~$\lesssim M_{\mu'}\lesssim$~ 340~(350)~GeV.
In contrast, the right plot shows that the complex $S$ boson is preferred to have a small mass for $S$.  Explicitly, we find 7~(7) GeV~$\lesssim m_S\lesssim$~14~(14) GeV while 130~(100)~GeV~$\lesssim M_{\mu'}\lesssim$~240~(400)~GeV.}  Such different behaviors in $m_S$ between the two scenarios are rooted in the $d$- and $p$-wave scattering cross sections given in Eqs.~\eqref{eq:anni-rl} and \eqref{eq:anni-cplx}.

\begin{figure}[tb]
\begin{center}
\includegraphics[width=80mm]{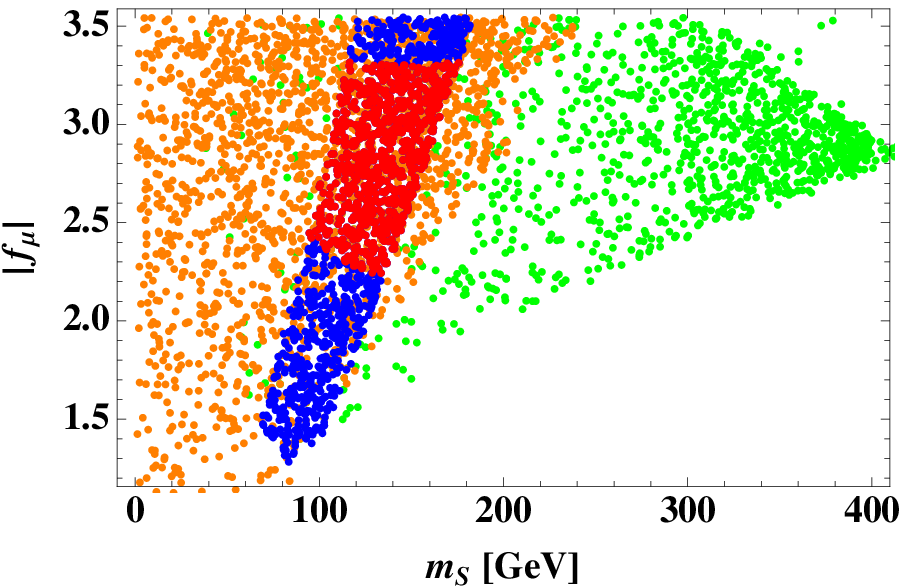}
\includegraphics[width=80mm]{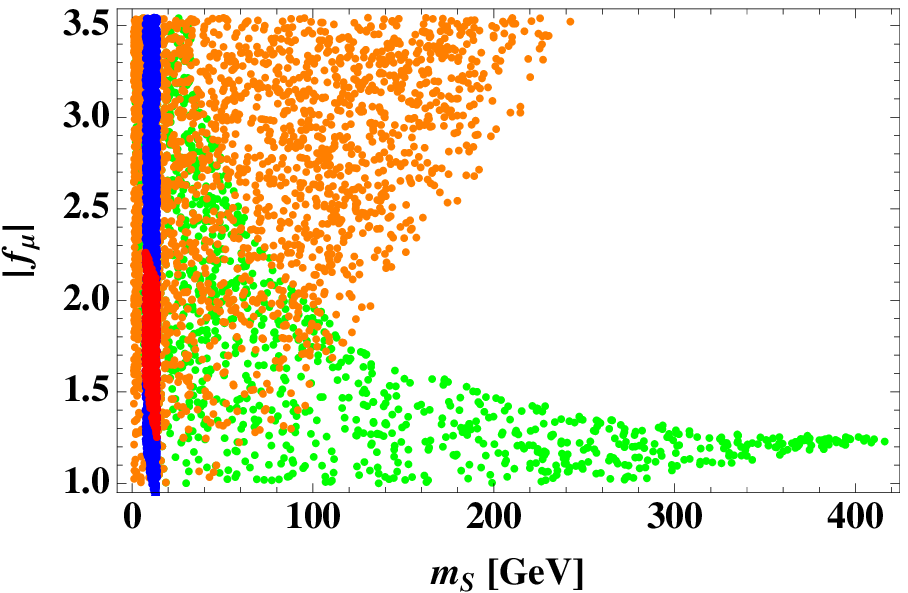}
\caption{Same as Fig.~\ref{fig:dm-ME}, but {in} the $(m_S,|f_\mu|)$ plane.}
\label{fig:dm-fmu}
\end{center}
\end{figure}

In Fig.~\ref{fig:dm-fmu}, we show scatter plots in the $(m_S,|f_\mu|)$ plane in the same style as in Fig.~\ref{fig:dm-ME}. 
The left plot shows that the allowed range of $|f_\mu|$ is {${2.2~(1.3)}\lesssim |f_\mu|\lesssim3.3~(\sqrt{4\pi})$, while the right plot has $1.2~(0.9)\lesssim |f_\mu|\lesssim2.3~(\sqrt{4\pi})$}, with $\sqrt{4\pi}$ being the limit of perturbativity.

\section{Lepton Flavor-Violating Processes}
\label{sec:Zll}

Our model can be extended to have three families of $L'$: $L^T_i\equiv [N'_{e,\mu,\tau},L'_{e,\mu,\tau}]^T$ so that each of the SM lepton family has corresponding exotic leptons.  In this case, we have to consider lepton flavor-violating (LFV) processes at the one-loop level, as they can arise from the mixing between the electron and tau flavor eigenstates in this model.  Here $L'_{e,\tau}$ have the same charges of $L'_{\mu}$ except that they are neutral under the $U(1)_\mu$ symmetry.
{Thus, $L'_\mu$ does not mix with the other flavors $L'_{e,\tau}$; therefore $L'_\mu$ is regarded as the mass eigenstate.}
First, the mass matrix of the exotic charged lepton masses is given by
\begin{align}
M_{L'}= 
\begin{pmatrix}
M_{e} & M_{e\tau}
\\
M_{\tau e} & M_{\tau}
\end{pmatrix}
~.
\end{align}
The mass eigenvalues are obtained through a bi-unitary transformation on the left-handed and right-handed fields: {${\rm diag}(M_{\ell'_e},M_{\ell'_\tau})=V^\dag_{\ell'_L} M_{L'} V_{\ell'_R}$.
Therefore, ${\rm diag}(|M_{\ell'_e}|^2,|M_{\ell'_\tau}|^2) \equiv V^\dag_{\ell'_L} M_{L'} M_{L'}^\dag V_{\ell'_L}$.}
Finally we find the following relation between the flavor and mass eigenstates:
\begin{align}
\begin{pmatrix}
\ell'_{e} \\
\ell'_{\tau}
\end{pmatrix}_f=
\begin{pmatrix}
c_\alpha & -s_\alpha\\
s_\alpha& c_\alpha
\end{pmatrix}
\begin{pmatrix}
\ell'_{e} \\
\ell'_{\tau} 
\end{pmatrix}_m
~,
\end{align}
where $s_\alpha \equiv \sin\alpha$ and $c_\alpha \equiv \cos\alpha$.
In the following discussions, we will always refer to the mass eigenstates.
In the mass eigenbasis, the relevant interactions are
\begin{align}
f'_{ee} \bar \ell_e P_R \ell'_e S +  f'_{e\tau} \bar \ell_e P_R \ell'_\tau S
+f'_{\tau e} \bar \ell_\tau P_R \ell'_e S +  f'_{\tau\tau} \bar \ell_\tau P_R \ell'_\tau S+{\rm H.c.}
~,
\end{align}
where $f'_{ee} = s_\alpha f_{e\tau}+c_\alpha f_{e}$, $f'_{e\tau} = c_\alpha f_{e\tau}-s_\alpha f_e$, $f'_{\tau e} = c_\alpha f_{\tau e}+s_\alpha f_\tau$, and $f'_{\tau\tau} = c_\alpha f_{\tau}-s_\alpha f_{\tau e}$.

{\it Two-body decays}: Because of the feature that the mixing does not involve the muon,
we here consider the constraints from the $\tau\to e\gamma$ and $Z \to \tau e$ decays.  The relevant branching ratio formulas for the two modes can be lifted from Ref.~\cite{Chiang:2017tai}.  First, we have
\begin{align}
&
{\rm BR}(\tau\to e\gamma) 
\nonumber \\
&\approx \frac{0.1784\alpha_{em} }{{768}\pi G_F^2}
\left|
\sum_{a=e,\tau} f'_{ea} f'^\dag_{a\tau} 
{\frac{2m_S^6+3m_S^4 M_{\ell'_a}^2-6 m_S^2 M_{\ell'_a}^4+M_{\ell'_a}^6
+6 m_S^4 M_{\ell'_a}^2\ln \frac{M_{\ell'_a}^2}{m_S^2} }{(m_S^2 - M_{\ell'_a}^2)^4}}
\right|^2
~.
\label{eq:teg}
\end{align}

We also obtain
\begin{align}
&\text{BR}(Z\to\tau e)
=
\frac{G_F}{3\sqrt2 \pi} \frac{m_Z^3 }{(16\pi^2)^2 \Gamma_{Z}^{\rm tot}}\left(s_w^2-\frac12\right)^2
\left|\sum_{a=e,\tau} f'_{ea} f'^\dag_{a\tau} \left[ F_2(\ell'_i,S)+F_3(\ell'_i,S) \right]\right|^2
~,
\label{eq:Zll}
\end{align}
with $\Delta\equiv -xy m_Z^2+(x+y)(M_a^2-m_b^2)+m_b^2$ and the total $Z$ decay width $\Gamma_{Z}^{\rm tot} = (2.4952 \pm 0.0023)$~GeV~\cite{pdg}.
It is noted that the combination $\sum_{a=e,\tau} f'_{ea} f'^\dag_{a\tau}$ appear in both Eqs.~(\ref{eq:teg}) and (\ref{eq:Zll}), showing the correlation between the two observables in this model.
The current upper bounds on ${\rm BR}(\tau\to e\gamma)$ and ${\rm BR}(Z\to\tau e)$ are found to be~\cite{pdg}:
\begin{align}
{\rm BR}(\tau\to e\gamma) \lesssim 3.3\times10^{-8}
~~\mbox{and}~~
{\rm BR}(Z\to\tau e) < 9.8\times10^{-6}
\label{eq:zmt-exp}
\end{align}
at 90 \% CL and 95 \% CL, respectively.

{\it Three-body decays}:
In our case, we consider the $\tau\to e\mu\bar\mu$ decay due to the muon-specific interaction structure.\
\footnote{The constraint from the $\tau \to e e \bar e$ decay is weaker.}
In the approximation of heavy exotic leptons, the effective Hamiltonian for the decay is obtained from a box diagram to be
\begin{align}
{\cal H}_{\rm eff}(\tau\to e\mu\bar\mu)
&=\frac{|f_\mu|^2}{(4\pi)^2 }
\sum_{i=e,\tau}{(f'_{ei} f'^{\dag}_{i\tau})
G_{box}(m_{S},M_{\ell'_{i}},M_{\mu'}})(\bar\ell_\tau \gamma^\rho P_L\ell_e)(\bar\ell_\mu \gamma_\rho P_L\ell_\mu)+{\rm c.c.}
\nn\\
&\equiv
{C_{\tau\to e\mu\bar\mu}}
(\bar\ell_\tau \gamma^\rho P_L\ell_e)(\bar\ell_\mu \gamma_\rho P_L\ell_\mu)+{\rm H.c.}~,
\end{align}
where $C^{-1}_{\tau\to e\mu\bar\mu}$ has the dimension of mass squared. 
The branching ratio is then found to be~\cite{Crivellin:2013hpa} 
\begin{align}
{\rm BR}(\tau\to e\mu\bar\mu)\approx \frac{m^5_{\tau}}{1526 \pi^3 \Gamma_{\tau}}
\left|{C_{\tau\to e\mu\bar\mu}}\right|^2
~,
\end{align} 
where  $\alpha_{em}\approx{1/134}$~\cite{Jens-Erler} is the fine structure constant at the $m_\tau$ scale, $\Gamma_{\tau}\approx2.27\times 10^{-12}$~GeV is the total decay rate of the tau lepton, and should be smaller than the upper bound of $2.7\times 10^{-8}$ at the 90\% CL~\cite{pdg}.

\begin{figure}[tb]
\begin{center}
\includegraphics[width=90mm]{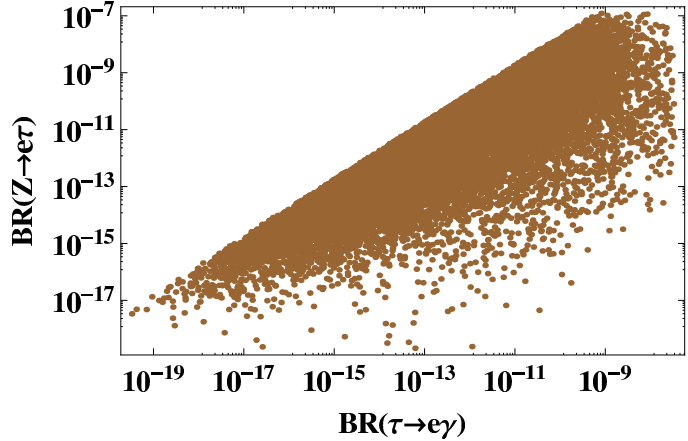}
\caption{Predicted distributions of ${\rm BR}(\tau\to e\gamma)$ and ${\rm BR}(Z\to e\tau)$ based upon the global scan.  
}
\label{fig:lfvs}
\end{center}
\end{figure}

In Fig.~\ref{fig:lfvs}, we show the distribution of ${\rm BR}(\tau\to e\gamma)$ and ${\rm BR}(Z\to e\tau)$ according to our global scan,
{where we restrict the coupling $f' \in [0.001,1]$}.  Note that this result is independent of whether the $S$ boson is real or complex.
The upper bound on ${\rm BR}(Z\to e\tau)$ comes from the same structure of Yukawa combination $(f'f'^\dag)_{e\tau}$ as shown in Eqs.~(\ref{eq:teg}) and~(\ref{eq:Zll}).
Since both of ${\rm BR}(\tau\to e\gamma)$ and  ${\rm BR}(Z\to e\tau)$ can reach up to their current experimental bounds, they can be tested in near future.

\section{Extension to quark sector \label{sec:extension}}

In view of the recent anomalies in $B$ physics, we extend the model to have one family of vector-like exotic quarks $Q'$ that are $SU(2)_L$ doublet.  However, the $S$ field has to be complex\
\footnote{With a real singlet $S$, it is impossible to explain the $b\to s\ell^+\ell^-$ anomalies because of a cancellation between diagrams~\cite{Arnan:2016cpy}.}.
Note also that one is not allowed to introduce an additional global symmetry similar to $U(1)_\mu$ above because the quark mixing or quark masses cannot be reproduced.  The relevant Lagrangian for the quark sector is then given by
\begin{align}
-{\cal L}_Q =
y_{u_{ij}} \bar Q_i \tilde H P_R u_{R_j} + y_{d_{ij}} \bar Q_i  H P_R d_{j} 
+ g_{i} \bar Q_i  P_R Q' S
+ M_{Q'_{} } \bar Q' Q'
+{\rm H.c.}
~,
\label{eq:quark}
\end{align}
where $Q'\equiv [u',d']^T$, $i,j \in \{ u, c, t \}~ (\{ d, s, b \})$ for the up-quark (down-quark) sector.
The first two terms are the same as the ones in the SM, while the third term is a new interaction that is important to the phenomenology discussions below. 
For the subsequent discussions, the relevant interactions in the mass eigenbasis are:
\begin{align}
f_\mu \bar \ell_\mu P_R \ell'_\mu S + g_{i} (\bar u_i  P_R u' +\bar d_i  P_R d') S + {\rm H.c.} ~.
\label{eq:qlop}
\end{align}

\subsection{$B\to K^{*} \bar\ell\ell$ anomalies}

First, the effective Hamiltonian for the $b \to s \mu^+ \mu^-$ transition induced by the operators in Eq.~\eqref{eq:qlop} through box diagrams\
\footnote{Although there exist penguin diagrams, they are subdominant because of the strong constraint from the $b\to s\gamma$ decay~\cite{Lees:2012ufa}.}
is~\cite{Arnan:2016cpy}
\begin{align}
&
{\cal H}_{\rm eff}(b\to s\mu^+ \mu^-) =
\frac{(g_{s} g^{\dag}_{ b}) |f_{\mu}|^2 }{(4\pi)^2}  
G_{box}(m_{S},M_{Q'}, M_{\mu'})
(\bar s \gamma^\rho P_L b)(\bar\ell_\mu\gamma_\rho \ell_\mu - \bar\ell_\mu\gamma_\rho \gamma_5\ell_\mu)
+{\rm H.c.}
\label{eq:btosmumu}
\nn\\
&\hspace{1cm}\equiv -C_{SM}\left[{C_9^{\mu\mu}} (O_9)_{\mu\mu} - {C_{10}^{\mu\mu}}  (O_{10})_{\mu\mu}\right]+{\rm H.c.}
~,
\\
& \mbox{with}~
G_{box}(m_{S},M_{Q'}, M_{\mu'})
\approx
\frac12\int_0^1dx_1\int_0^{1-x_1}dx_2
\frac{x_1}
{x_1 m^2_{S} +x_2 M^2_{Q'} +(1-x_1-x_2) M^2_{\mu'} },
\nn
\end{align}
 where $C_{SM}\equiv \frac{V_{tb} V^*_{ts}G_F\alpha_{em}}{\sqrt2\pi}$, $V_{tb}\sim0.9991$ and $V_{ts}\sim-0.0403$ are the Cabibbo-Kobayashi-Maskawa (CKM) matrix elements~\cite{pdg}.
Remarkably, we have $C_9^{\mu\mu}= -C_{10}^{\mu\mu}$ for the new physics contribution, which is one of the preferred schemes to explain the $B$ anomalies~\cite{Descotes-Genon:2015uva} and its $1\sigma$ ($3\sigma$) range is given by $[-0.85,-0.50]$ $([-1.22,-0.18])$ with the best fit value at $-0.68$.

Moreover, the interactions in Eq.~\eqref{eq:qlop} also lead to
\begin{align}
&
{\mathcal H}_{\rm eff}^{b d_\alpha} =
\frac{(g_{d_\alpha } g^{\dag}_{ b}) |f_{\mu}|^2 }{(4\pi)^2}  
G_{box}(m_{S},M_{Q'}, M_{\mu'})
(\bar d_\alpha \gamma_\mu P_L b)(\bar\ell_\mu\gamma^\mu P_L \ell_\mu)+{\rm H.c.}
\nn\\
&\hspace{1cm}\equiv
  - C_{LL}^{\bar d_\alpha b} 
(\bar d_\alpha \gamma_\mu P_L b)(\bar\ell_\mu\gamma^\mu P_L \ell_\mu)+{\rm H.c.}
\end{align}
for $d_\alpha \in \{ d, s \}$.  Therefore, the parameters have to satisfy the constraints from the data or upper bound of ${\rm BR}(B_{d/s}\to\mu^+\mu^-)$ reported by CMS~\cite{Chatrchyan:2013bka} and LHCb~\cite{Aaij:2013aka}.
The bounds on the coefficients $|C_{LL}^{\bar sb(\bar db)}|$ in the above effective Hamiltonian are given by~\cite{Sahoo:2015wya}
\begin{align}
|C_{LL}^{\bar sb(\bar db)}|\lesssim5~(3.9)\times 10^{-9}\ {\rm GeV}^{-2}
~.
\label{eq:Bdmumu}
\end{align}

\subsection{Neutral meson mixing}

The operators in Eq.~\eqref{eq:qlop} also contribute to neutral meson mixing at low energies.  Therefore, the couplings $g'$ and masses $M_{Q'}$ are strongly constrained by the measured data.  It is straightforward to obtain the appropriate effective Hamiltonian for meson mixing by the replacements $f'\to g$ and $M_{\mu'} \to M_{Q'}$ in Eq.~\eqref{eq:btosmumu}. 
The mass splitting between neutral mesons $M$ and $\bar M$ is then
\begin{align}  
 \Delta M_M&=
 \frac{2m_M f_M^2}{3(4\pi)^2} 
 (g_{\beta } g^{\dag}_{\alpha})(g_{b} g^{\dag}_{a})
 G_{box}(m_{S},M_{Q'}, M_{Q'})
 ~.
\end{align}
Here we take into account the $K^0-\bar K^0$, $B_d-\bar B_d$, and $B_s-\bar B_s$ mixings\
\footnote{The constraint from $\Delta m_D$ is weaker than that of $\Delta m_K$.}
~\cite{Gabbiani:1996hi}:
\begin{align}
& \Delta m_K :
(g_{2} g^{\dag}_{1})(g_{1} g^{\dag}_{2})  G_{box} \lesssim 3.48\times10^{-15}
{\times \frac{24\pi^2}{m_K f_K^2}}  {{\rm GeV}}
~,
\label{eq:kk}\\
& \Delta m_{B_d} :
(g_{3 } g^{\dag}_{1})(g_{1} g^{\dag}_{3})  G_{box} \lesssim 3.3{4}\times10^{-13}
{\times \frac{24\pi^2}{m_{B_d} f_{B_d}^2}}{{\rm GeV}}
~,\\
& \Delta m_{B_s} :
(g_{3 } g^{\dag}_{2})(g_{2} g^{\dag}_{3})  G_{box} \lesssim 1.17\times10^{-11}
{\times \frac{24\pi^2}{m_{B_s} f_{B_s}^2}} {{\rm GeV}}
~,
\end{align}
where $ G_{box} \equiv G_{box}(m_{S},M_{Q'}, M_{Q'})$.  The other parameters are also found to be
$f_K = 0.160$~GeV, $f_{B_d(B_s)} = 0.200$~GeV~\cite{Gabbiani:1996hi}, 
$m_K = 0.498$~GeV, and $m_{B_d(B_s)} = 5.280~ (5.367)$~GeV~\cite{pdg}. 
One finds that these constraints are not generally so stringent.
When $g_{i}=0.1$ is taken universally, for example,
all the bounds are always satisfied with the most stringent bound coming from $\Delta m_K$.

\begin{figure}[tb]
\begin{center}
\includegraphics[width=90mm]{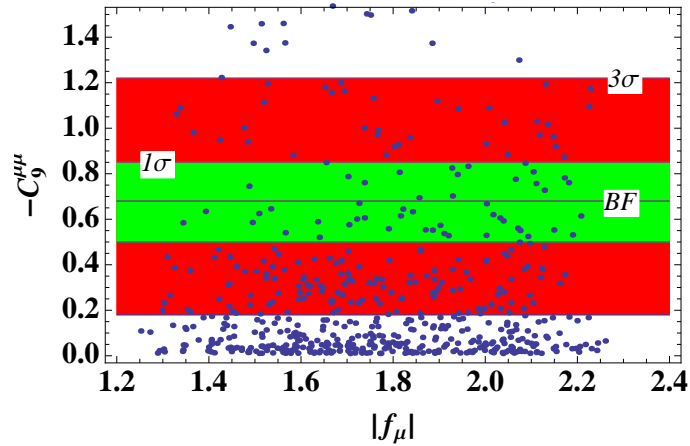}
\caption{{Global analysis in the plane of $|f_\mu|$ and $-C_9^{\mu\mu}$, satisfying all the constraints discussed in the main text.
Note here that only the complex $S$ allows a non-vanishing $-C_9^{\mu\mu}$.
The black horizontal line at the center of the green region represents the best-fit value for the observed anomaly in $b\to s \mu^+ \mu^-$ decays,
with the green (red) region being the $1\sigma$ ($3\sigma$) range of the current experimental bound~\cite{Descotes-Genon:2015uva}.
}}
\label{fig:fmu-c9}
\end{center}
\end{figure}

Here we analyze whether there is any parameter space in
the allowed region for a complex scalar
(7~GeV $\lesssim m_S\lesssim$ 14~GeV, {130~GeV $\lesssim M_{\mu'}\lesssim$ 240~GeV, and 1.2 $\lesssim |f_\mu|\lesssim 2.3$} found in the previous section) that can satisfy the $B$ anomalies.
In Fig.~\ref{fig:fmu-c9}, we show the a scatter plot in the $(|f_\mu|,-C_9^{\mu\mu})$ plane,
where we have selected the input parameters: $g \in [0.001,\sqrt{4\pi}]$ and $M_{Q'}\in [500,2000]$~GeV.
The central black horizontal line represents the best fit value of $- C_9^{\mu\mu}$, and
the green (red) region is the $1\sigma$ ($3\sigma$) range~\cite{Descotes-Genon:2015uva}.  
{We have found many parameter sets satisfying the required range of $-C_9^{\mu\mu}$.  They span the entire range of $|f_\mu|$ that can explain the muon anomalous magnetic moment, as we vary the other parameter $g$ in Eq.~(\ref{eq:btosmumu}) that is not strongly restricted by $B_{d/s}\to\mu^+\mu^-$ decays and neutral meson mixing.}
It is seen that through a simple extension to the quark sector in a way analogous to the lepton sector, the model can readily accommodate the $B$ anomalies as well.\\

{\subsection{Constraints from direct production of $Q'$s and $L'$s at LHC}

The exotic quark pair production can be induced via QCD processes at the LHC.
Each $Q'$ then decays via $Q' \rightarrow q_i S$,
where $q_i$ denotes some quark of flavor $i$.
Hence, searching for ``$\{ t t, b b , t  j, b j , j j  \}$ + missing $E_T$" signals
will impose stringent constraints on our model.
The branching ratio for a particular quark flavor $i$ depends on
the relative size of the Yukawa couplings, $g_{3j}$ and $g_{aj}$ with $a = 1, 2$.  
As a result, the lower limit on the mass of $Q'$ can be roughly estimated using
the current LHC data for the squark searches~\cite{CMS:2016mwj, Aaboud:2016zdn},
which indicates the mass should be larger than $\sim 0.5 - 1$~TeV
with details depending on the mass difference between $Q'$ and $S$.
This range is consistent with the selected parameter values in this work.
While $L'$ pair production can be generated via $Z/\gamma$ processes, its production rate is smaller than that of $Q'$.
Therefore, the constraint of mass on $L'$ will not be as stringent.
\footnote{If $L'$ behaves like sleptons~\cite{atlas}, the lower mass bound would be around 300~GeV, posing a difficulty to the $Z_3$ model. } 
}

\section{Conclusions \label{sec:summary}}

We have proposed a model with muon-specific interactions, with the intent to explain the muon anomalous magnetic dipole moment and the dark matter relic density.
In the model, we impose a global $U(1)_\mu\times {Z_N}$ symmetry, and introduce an exotic lepton iso-doublet, three right-handed Majorana fermions, and an inert scalar iso-singlet in addition to the SM field contents.  In the case of a real (complex) scalar boson, we take $N = 2$ ($N = 3$ as a simplest choice).  Leptons in the second family and the corresponding exotic lepton is charged under the $U(1)_\mu$ symmetry.  The exotic lepton doublet and the inert scalar field have nontrivial $Z_N$ charges.
{Note that here the $U(1)_\mu$ and $Z_{2,3}$ symmetries play an important role in the model construction and renders a minimal framework for accommodating all the recently observed muon-related anomalies.}

As a result of such an extension, the model features a good DM candidate and the capacity to accommodate $\Delta a_\mu$.  We have studied both scenarios of real and complex $S$ as the weakly interacting massive particle DM.  Through a comprehensive scan by also including constraints from lepton flavor violating processes, we have obtained the following allowed parameter space: 
{  \begin{align*}
&90~{\rm GeV}\lesssim m_S\lesssim 180~{\rm GeV},
\quad 
2.2\lesssim |f_\mu|\lesssim 3.3,
\quad 
160~{\rm GeV}\lesssim M_{\mu'}\lesssim 340~{\rm GeV},
\quad 
({\rm Real\ S})
,\\
&7~{\rm GeV}\lesssim m_S\lesssim 14~{\rm GeV},
\quad 
1.2\lesssim |f_\mu|\lesssim 2.3,
\quad 
130~{\rm GeV}\lesssim M_{\mu'}\lesssim 240~{\rm GeV},\quad ({\rm Complex\ S}).
\end{align*}
Here we have found that the constraint of BR$(Z\to\mu^+ \mu^-)$ restricts our parameter space.
}
{The fact that the DM masses obtained above are at the electroweak scale makes the scheme highly testable in various on-going experiments.  Moreover, our result points out a way to discriminate the two scenarios as they have distinctly different ranges for the DM mass.}

In view of the recent $B$ anomalies, we have also extended the model to the quark sector in a way analogous to the lepton sector, except that no additional global symmetry needs to be introduced.  We have found that the preferred Wilson coefficients in the effective Hamiltonian of the $b \to s \mu^+ \mu^-$ transitions can be readily obtained while being consistent with constraints from, for example, the $B_{s(d)}\to \mu^+\mu^-$ decay and neutral meson mixings.  In particular, the scenario of a real $S$ cannot explain the anomalies due to a cancellation between new physics contributions.  The scalar therefore has to be complex and has a mass of ${\cal O}(10)$~GeV,
{a testable observable of the scheme.}

\section*{Acknowledgments}

The authors thank Mr. Guan-Jie Huang for his participation at the early stage of the research.
This work was support in part by the Ministry of Science and Technology of ROC under Grant No.~MOST~104-2628-M-002-014-MY4.
This research is supported by the Ministry of Science, ICT and Future Planning, Gyeongsangbuk-do and Pohang City (H.O.).
H. O. is sincerely grateful for KIAS and all the members.


\end{document}